# Exact solution of the Schrödinger equation for the inverse square root potential $V_0/\sqrt{x}$

**A.M. Ishkhanyan**

Institute for Physical Research, NAS of Armenia, 0203 Ashtarak, Armenia

**Abstract.** We present the exact solution of the stationary Schrödinger equation for the potential $V = V_0/\sqrt{x}$. Each of the two fundamental solutions that compose the general solution of the problem is given by a combination with non-constant coefficients of two confluent hypergeometric functions of a shifted argument. Alternatively, the solution is written through the first derivative of a tri-confluent Heun function. Apart from the quasi-polynomial solutions provided by the energy specification $E_n = E_1 n^{-2/3}$, we discuss the bound-state wave functions vanishing both at infinity and in the origin. The exact spectrum equation involves two Hermite functions of non-integer order which are not polynomials. An accurate approximation for the spectrum providing a relative error less than $10^{-3}$ is $E_n = E_1 (n - 1/(2\pi))^{-2/3}$. Each of the wave functions of bound states in general involves a combination with non-constant coefficients of two confluent hypergeometric and two non-integer order Hermite functions of a scaled and shifted coordinate.

## 1. Introduction

The exact solutions of the Schrödinger equation are rare. The most known ones are the potentials solvable in terms of the ordinary and confluent hypergeometric functions. The list of the known ordinary hypergeometric potentials is rather large; however, there exist only two independent such potentials: the Eckart [1] and the Pöschl-Teller [2] potentials. Many of the widely used potentials such as the Rosen-Morse [3], Woods-Saxon [4], Manning-Rosen [5], Hulthén [6], Scarf [7] ones, are in fact particular specifications of the mentioned two independent potentials. In the confluent hypergeometric case there are three independent potentials; the harmonic oscillator (plus inverse square) potential discussed by Schrödinger [8], the Kratzer potential (Coulomb plus inverse square) [8,9] and the Morse potential [10].

In the present paper we present one more potential solvable in terms of the confluent hypergeometric functions - the inverse square root potential $V = V_0/\sqrt{x}$. The exact solution of the Schrödinger equation for this potential has a remarkable structure. Each of the two fundamental solutions that compose the general solution of the problem presents a combination of two confluent hypergeometric functions. The coefficients of the combinations are not constants. It is convenient to write one of the confluent hypergeometric functions involved in the fundamental solutions as a Hermite function of non-integer order which, in general, is not a polynomial. Then the standard polynomial reduction is achieved by the energy specification $E_n = E_1 n^{-2/3}$. Furthermore, we discuss the bound-state wave functions vanishing both at infinity and in the origin. The exact energy spectrum is derived through an equation involving two Hermite functions. We show that a highly accurate approximation for the energy levels is given as $E_n = E_1(n - 1/(2\pi))^{-2/3}$. This approximation provides an absolute error less than $10^{-3}$ for all quantum numbers $n \in N$.

We note that the general solution of the problem is derived via the reduction of the Schrödinger equation to an equation obeyed by a function involving the first derivative of a tri-confluent Heun function [11-13]. The resultant two-term confluent hypergeometric fundamental solution for the inverse square root potential is in fact a specification of the derivative of a tri-confluent Heun function for given parameters. For this reason, the derivation is itself of methodological interest because it introduces a different feature in the search for solvable potentials. For this reason, the derivation is separated out to an appendix.

## 2. General solution and quasi-polynomial reductions

It is straightforwardly checked by direct substitution that the general solution of the one-dimensional stationary Schrödinger equation for a particle of mass $m$ and energy $E$:

$$\frac{d^2\psi}{dx^2} + \frac{2m}{\hbar^2}(E - V(x))\psi = 0, \qquad (1)$$

for the inverse square root potential

$$V = \frac{V_0}{\sqrt{x}} \qquad (2)$$

with arbitrary $V_0$ is written as

$$\psi(x) = e^{-\delta x/2}\frac{du}{dy}, \qquad (3)$$

where

$$u = e^{-\sqrt{2a}\,y}\left(c_1 \cdot H_a(y) + c_2 \cdot {}_1F_1\left(-\frac{a}{2};\frac{1}{2};y^2\right)\right). \qquad (4)$$

Here $c_{1,2}$ are arbitrary constants, $H_a$ is the Hermite function [14] (for a non-negative integer $a$ it becomes the Hermite polynomial; however, in general $a$ is arbitrary), ${}_1F_1$ is the Kummer confluent hypergeometric function [14], the auxiliary dimensionless argument $y$ defines a scaling of the coordinate followed by deformation and shift:

$$y = \text{sgn}(V_0)\sqrt{\delta}x + \sqrt{2a}, \tag{5}$$

and the involved parameters $\delta$ and $a$ are given as

$$\delta = \sqrt{-8mE/\hbar^2}, \tag{6}$$

$$a = \frac{m^2 V_0^2}{\hbar(-2mE)^{3/2}}. \tag{7}$$

A standard set of bounded quasi-polynomial solutions for an attractive potential with $V_0 < 0$ is achieved by putting $a = n$, $n \in N$. Then, the Hermite function in equation (4) becomes the Hermite polynomial and one should put $c_2 = 0$ to ensure vanishing of the solution at infinity. The energy eigenvalues for this case are

$$E_n = \frac{V_0}{2}\left(\frac{-mV_0}{\hbar^2}\right)^{1/3} n^{-2/3}, \quad n = 1, 2, 3, \ldots, \tag{8}$$

and the corresponding solutions are written as

$$\psi_n = e^{-\sqrt{2n}\,y - \delta x/2}(H_n(y) - \sqrt{2n}\,H_{n-1}(y)), \quad y = \sqrt{2n} - \sqrt{\delta}x. \tag{9}$$

To get an explicit representation, here are the first three quasi-polynomials:

$$\psi_1 = e^{-\sqrt{2}\,y - \delta x/2}(1 - \sqrt{2}\,y), \tag{10}$$

$$\psi_2 = e^{-2y - \delta x/2}(1 + 2y - 2y^2), \tag{11}$$

$$\psi_3 = e^{-\sqrt{6}\,y - \delta x/2}(3 - 3\sqrt{6}\,y - 6y^2 + 2\sqrt{6}\,y^3). \tag{12}$$

A peculiarity of this set of quasi-polynomial functions is that the solutions do not vanish in the origin (Fig.1). However, depending on the particular physical problem at hand (for instance, if one considers equation (1) as the s-wave radial equation for the three-dimensional Schrödinger equation with the potential $V = V_0/\sqrt{r}$), it is useful to have a set of bound wave functions that vanish at the origin [15].

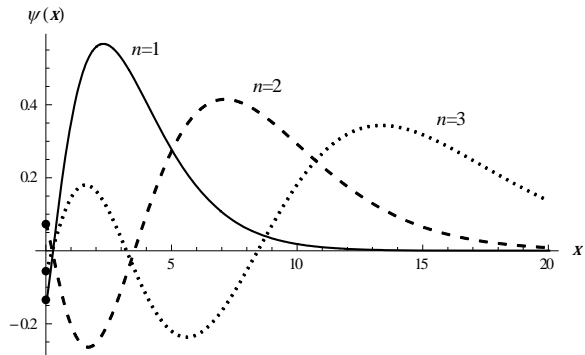

Fig. 1. The first three bounded quasi-polynomial solutions for the energy specification (8), $V_0 = -1$, $\hbar = m = 1$.

## 3. Bound states

Consider the case when one demands the wave function to vanish both in the origin and at the infinity: $\psi(0)=0$ and $\psi(+\infty)=0$ ($V_0<0$). The first condition imposes the following relation between the coefficients $c_{1,2}$:

$$c_1 = \frac{2a_1 F_1\left(1-\frac{a}{2};\frac{3}{2};2a\right) + {}_1F_1\left(-\frac{a}{2};\frac{1}{2};2a\right)}{\sqrt{2a}\, H_{a-1}\left(\sqrt{2a}\right) - H_a\left(\sqrt{2a}\right)} c_2, \tag{13}$$

while the second one, after matching the leading asymptotes of the involved Hermite and confluent hypergeometric functions, results in a transcendental equation for the parameter $a$, which is conveniently written through two Hermite functions as

$$\sqrt{2a}\, H_{a-1}\left(-\sqrt{2a}\right) + H_a\left(-\sqrt{2a}\right) = 0. \tag{14}$$

This is the equation for the energy spectrum. It possesses a countable infinite set of roots which, together with equation (7), determine the energy eigenvalues. To solve this equation, we divide it by $H_a(-\sqrt{2a})$ and use the known recurrence relations and series expansions for the involved Hermite functions [14] to get the following key approximation

$$F \equiv \frac{\sqrt{2a}\, H_{a-1}(-\sqrt{2a})}{H_a(-\sqrt{2a})} + 1 \approx \frac{\sin\left(\pi a + 1/2 - a e^{-2a}\right)}{\sqrt{a}\left(1+\log(1-b_1/\sqrt{2a})\right)\cos\left(\pi a + 1 + \log(1-b_2/\sqrt{2a})\right)}, \tag{15}$$

where the parameters $b_{1,2}$ are constants of the order of unity. The accuracy of this result is demonstrated in Fig. 2, where the filled circles indicate the exact numerical values and the solid curve presents the derived approximation.

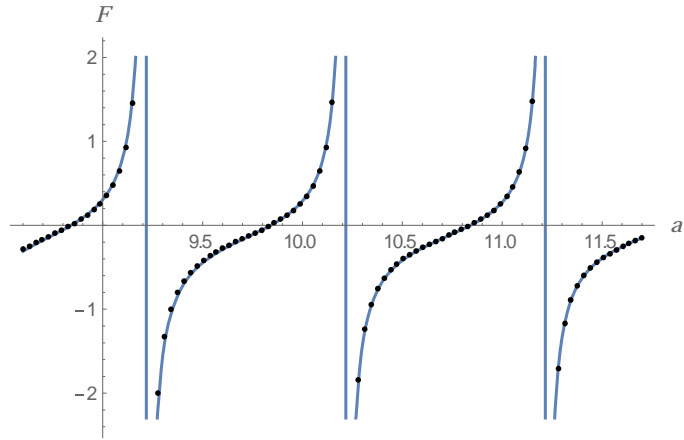

Fig. 2. Approximation (10) - solid curve, exact numerical values - filled circles.

Having established this approximation, we arrive at a remarkably simple, yet highly accurate, approximation for the exact eigenvalue equation (14) written in terms of the familiar elementary functions:

$$\sin\left(\pi a + 1/2 - a e^{-2a}\right) = 0. \tag{16}$$

Compared with the exact equation, everywhere for $a > 1/2$ this approximation provides a relative error of the order or less than $10^{-3}$.

The smallest root of equation (16) is close to unity ($a_1 \approx 0.86$); consequently, the exponential term is always small. Hence, even if this term is neglected, one may expect a sufficiently good approximation for the energy spectrum. Thus, the roots of equation (16) are approximately given as $a_n = n - 1/(2\pi)$. Equation (7) then gives the spectrum

$$E_n = \frac{V_0}{2}\left(\frac{-mV_0}{\hbar^2}\right)^{1/3}\left(n - \frac{1}{2\pi}\right)^{-2/3}, \quad n = 1, 2, 3, \ldots \tag{17}$$

This is indeed a rather accurate approximation. The relative error is less than $10^{-3}$ for all $n > 2$ and it is less than $10^{-5}$ for $n \geq 7$ (Fig. 3).

We note that since the roots $a_n$ are not integers the wave functions of the bound states for spectrum (17) are not quasi-polynomials in contrast to the ordinary spectrum (8). The wave functions of the ground state $n = 1$ and the next two states are shown in Fig. 4.

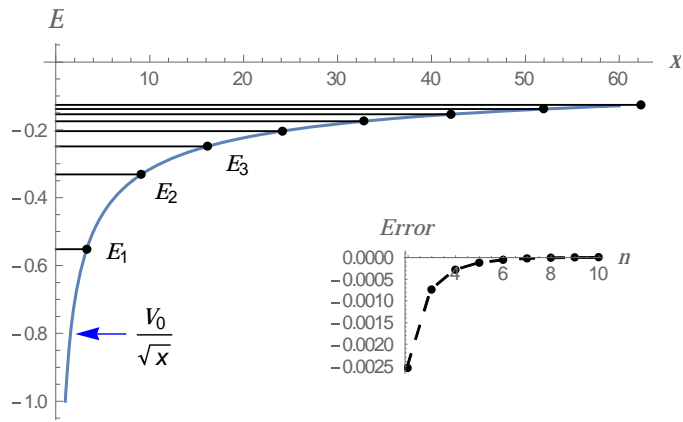

Fig. 3. Potential (2) and energy spectrum (17) for $V_0 = -1$, $\hbar = m = 1$. The inset shows the relative error of the $(n - 1/(2\pi))^{-2/3}$ approximation for the spectrum.

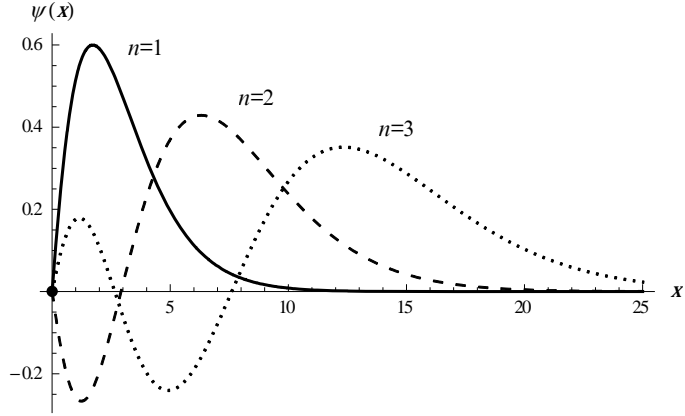

Fig. 4. Normalized wave functions of the ground state $n=1$ and the next two states for the energy spectrum (17), $V_0 = -1$, $\hbar = m = 1$.

## 4. Discussion

Thus, we have presented the general solution of the one-dimensional stationary Schrödinger equation for the inverse square root potential. The solution is written in terms of the confluent hypergeometric functions and the Hermite functions of non-integer order. We have shown that the standard quasi-polynomial solutions provide the spectrum $E_n = E_1 n^{-2/3}$.

Discussing the case of the bound-state wave functions vanishing both at infinity and in the origin, we have shown that the exact spectrum equation for this case involves two Hermite functions of non-integer order. We have derived an accurate approximation for the energy spectrum providing an absolute error less than $10^{-3}$ as $E_n = E_1 (n - 1/(2\pi))^{-2/3}$. Each of the bound-state wave functions involves a combination with non-constant coefficients of four confluent hypergeometric functions (or, equivalently, non-integer order Hermite functions) of a scaled and shifted coordinate argument. A distinct feature of these wave functions is that they are not quasi-polynomials.

We have shown that for the inverse square root potential the solution of the Schrödinger equation is equivalently written through the derivative of the solution of a tri-confluent Heun function. The employment of the equations obeyed by the derivatives of the Heun functions for construction of exactly solvable potentials seems to be promising. This is because, on the one hand, these equations are more complicated than the Heun equations (in that they have extra singularities) and, hence, they suggest a wider set of effects, and, on the other hand, their solutions are written through the solutions of the Heun equations and thus these functions do not cause additional complications compared to the Heun functions.

It is worth mentioning that the inverse square root potential is a member of one of the five exactly solvable six-parametric *bi-confluent* Heun potentials presented by Lamieux and Bose [16], see also [17]. (Note that our potential is actually three-parametric; apart from the parameter $V_0$ there are two more available parameters - the space origin and the energy origin that we have omitted.) Our result thus shows that there exists a particular specification of the parameters of the bi-confluent Heun equation for which the bi-confluent Heun function is expressed through the tri-confluent Heun functions. There are other cases supporting this observation, e.g., several quantum two-state models solvable in terms of the Heun functions suggest similar conclusions [18-20]. This applies not only to the bi-confluent Heun function but to all other Heun functions including the general Heun function [20].

The inverse square root potential appears in many studies as a part of a more general potential (see, e.g., [15,21,22]). However, these are either conditionally solvable potentials when a parameter of the potential is fixed or energy dependent potentials. We would like to emphasize that the inverse square root potential we have presented is an independent exactly solvable one without restrictions so that the solution is not affected by additional influences coming from extra terms involved in the potential.

A last remark concerns the Darboux transformation [23] which is widely applied to construct new solvable potentials by employing the known ones [24]. Though the solution that we have presented involves a special function and its derivative, the potential is not derived by a Darboux transformation. The latter assumes that the involved special function is a solution of a (different) Schrödinger equation, while in our case this is not the case. This assertion is further supported by the observation that because of a logarithmic derivative involved in the Darboux transformation it usually results in a conditionally integrable potential with a fixed parameter [24], while we present an exactly solvable one.

**Acknowledgments**

This research has been conducted within the scope of the International Associated Laboratory IRMAS (CNRS-France & SCS-Armenia). The work has been supported by the Armenian State Committee of Science (SCS Grant No. 13RB-052).**Appendix: derivation of the solution**

The derivatives of the Heun functions generally obey more complicated equations involving additional apparent singularities. For the tri-confluent Heun equation [11,12]

$$\frac{d^2u}{dz^2}+\left(\gamma+\delta z+\varepsilon z^2\right)\frac{du}{dz}+(\alpha z-q)u=0, \tag{A1}$$

if $\alpha$ is not zero, the equation obeyed by the function

$$w=e^{\gamma z+\delta z^2/2+\varepsilon z^3/3}\frac{du}{dz} \tag{A2}$$

is written as

$$\frac{d^2w}{dz^2}-\left(\gamma+\delta z+\varepsilon z^2+\frac{1}{z-z_0}\right)\frac{dw}{dz}+\alpha(z-z_0)w=0, \tag{A3}$$

where $z_0=q/\alpha$. It is seen that compared to the starting tri-confluent equation (A1) this equation possesses an additional singularity located at $z_0$.

We now try to reduce the Schrödinger equation to this equation by transforming both dependent and independent variables $\psi=\varphi(z)u(z)$, $z=z(x)$. The approach suggested in [17] for construction of exactly integrable energy-independent potentials is based on the assertion that if a potential is proportional to an energy-independent parameter, then the logarithmic $z$-derivative $\rho'(z)/\rho(z)$ of the function $\rho(z)=dz/dx$ cannot have poles other than the finite singularities of the target equation to which the Schrödinger equation is reduced. It then follows that if the target equation is an equation with rational coefficients, then the equation for the coordinate transformation should necessarily be of the form $\rho(z)=\Pi_i(z-z_i)^{A_i}$, where $z_i$ are the mentioned finite singularities of the target equation (Manning form [25]). In addition, the exponents $A_i$ should all be integers or half-integers.

Since the only finite singularity of equation (A3) is $z=z_0$, the only permissible coordinate transformation for this equation is given as $z'(x)=\rho=(z-z_0)^{m_1}/\sigma$ with integer or half-integer $m_1$. Following the lines of [17], we match the logarithmic derivative of $\rho(z)$ with the corresponding term of the invariant of equation (A3):

$$\frac{1}{2}\left(\frac{\rho_z}{\rho}\right)_z+\frac{1}{4}\left(\frac{\rho_z}{\rho}\right)^2=\frac{3/4}{(z-z_0)^2}. \tag{A4}$$

As a result we get $m_1=-1$. Consequently, the admissible coordinate transformation is

$$z^2=\frac{2(x-x_0)}{\sigma}. \tag{A5}$$

Now, we look for a solution of the Schrödinger equation of the form [17,18]

$$\psi=e^{-\left(\gamma z+\delta z^2/2+\varepsilon z^3/3\right)/2}w(z), \tag{A6}$$

where $w(z)$ is the solution of equation (A3). Substituting this $\psi$ and $z(x)$ from (A5) into the Schrödinger equation (1) and eliminating $w''(z)$ using equation (A3), we arrive at an equation which is proportional to $w(z)$. This is because the pre-factor in the wave function (A6) is chosen so that the term proportional to the first derivative $w'(z)$ is cancelled. The equation reads

$$\left( \frac{\sqrt{2}\gamma}{x^{3/2}} + \frac{\gamma^2}{x} - \frac{2\sqrt{2}(2\alpha - \gamma\delta + \varepsilon)}{x^{1/2}} + 2(\delta^2 + 2\gamma\varepsilon) + 4\sqrt{2}\delta\varepsilon x^{1/2} + 4\varepsilon^2 x \right.$$
$$\left. + \frac{16m(E-V)}{\hbar^2} \right) w(\sqrt{2x}) = 0, \tag{A7}$$

where for simplicity we have put $x_0 = 0$, $\sigma = 1$ and, without loss of the generality, $q = 0$. This specification of the parameters can always be applied because the tri-confluent Heun equation preserves its form when scaling and shifting the origin.

The last step is now straightforward. First, equation (A7) shows that $V(x)$ is necessarily of the form

$$V(x) = \frac{V_5}{x^{3/2}} + \frac{V_4}{x} + \frac{V_0}{x^{1/2}} + V_1 + V_2 x^{1/2} + V_3 x. \tag{A8}$$

Substituting this into equation (A7) and requiring the vanishing of the coefficients at powers of $x$, we readily get that if the parameters $V_{0,1,2,3,4,5}$ of the potential are supposed independent, then $V_5 = V_4 = V_3 = V_2 = 0$ and

$$\delta = \pm \frac{2\sqrt{2m(-E+V_1)}}{\hbar}, \quad \alpha = -\frac{2\sqrt{2}mV_0}{\hbar^2}, \quad \gamma = \varepsilon = 0. \tag{A9}$$

These parameters determine the tri-confluent Heun function that gives, through equations (A2),(A6), the solution of the Schrödinger equation for the inverse square root potential (2).

Notably, the parameter $\varepsilon$ for this solution is zero. It is known that then the tri-confluent Heun equation is reduced to the Kummer confluent hypergeometric equation via a transformation of the variables. This observation accomplishes the development. Because it is the derivative $w(z) = u'(z)$ of the tri-confluent function that is involved in the wave function (A6) and since the particular tri-confluent Heun function under consideration is a product of a confluent hypergeometric function and a non-constant pre-factor, we get a solution that generally involves four confluent hypergeometric functions (each fundamental solution of the corresponding confluent hypergeometric equation leads to a combination with non-constant coefficients of a pair of confluent hypergeometric functions).